\documentstyle[12pt]{article}

\begin{titlepage}
\title{{\bf Entropy for
dilatonic black hole }}
\author{Amit Ghosh and Parthasarathi Mitra
\date{June 1994\\hep-th/9406210}
\thanks{e-mail addresses amit, mitra@saha.ernet.in}\\
Saha Institute of Nuclear Physics\\
Block AF, Bidhannagar\\
Calcutta 700 064, INDIA}
\end{titlepage}
\begin{document}
\def\wt {\widetilde}
\maketitle
\abstract{
The area formula for  entropy is extended to the case of a dilatonic
black hole. The  entropy  of  a scalar field in
the background of such a black hole is calculated semiclassically.
The area and cutoff dependences are normal {\it except in the
extremal case}, where the area is zero but the entropy nonzero.
}
\newpage

It  is well known that the area of the horizon of a black
hole  can  be  interpreted as an entropy \cite{Bek} and satisfies
all  the  thermodynamical laws. This is not yet understood
in terms of the usual formulation of entropy as a measure of
the number of states available, but the na\"\i ve  Lagrangian  path
integral does lead to  a  partition  function  from  which the area
formula for entropy  can  be  obtained  \cite{GH}  by  neglecting
quantum fluctuations.

There have also been some attempts at calculating the entropy of quantum
fields  in  black  hole  backgrounds \cite{'tHooft, Uglum}. The values
thus obtained are contributions   to  the
entropy of the black hole -field system.
These calculations have produced
divergences,  but the area of the horizon has appeared as a factor.
This has been interpreted to mean that the gravitational
constant gets renormalized in the presence of the quantum fields \cite{Uglum}.
We shall investigate whether similar phenomena occur in the case
of  {\it dilatonic}  black  holes \cite{dil, review}
where it is possible to have a {\it vanishing} horizon area.

The   low- energy   limit   of   string   theory   with   unbroken
supersymmetry contains a massless dilaton field. Models
where  these  dilatons  are coupled with gravity may be used for
studying  black holes with small Compton  wavelengths.
The simplest four- dimensional model is
\begin{equation}
S={1\over 16\pi}\int d^4x\sqrt{-g}e^{-2\phi}(R+4g^{\mu\nu}\nabla_{\mu}\phi
\nabla_{\nu}\phi)
\end{equation}
where  $\phi$  is the massless dilaton field, $R$ is the
scalar  curvature   and   $g_{\mu\nu}$   is   the  metric. As the
curvature term contains an  extra  exponential  factor,  this  is
often removed by the conformal transformation
\begin{equation}
\wt g_{\mu\nu}=e^{-2\phi}g_{\mu\nu}.
\end{equation}
This changes the  action to
\begin{equation}
S={1\over 16\pi}\int d^4x\sqrt{-\wt g}(\wt R-2\wt g^{\mu\nu}\nabla_{\mu}\phi
\nabla_{\nu}\phi)\label{2},
\end{equation}
which  is standard Einstein gravity coupled with a massless scalar
field. Thus  $\wt g_{\mu\nu}$  is  the  appropriate  metric  for
gravitational  studies  and  in  fact  all  theorems  of  general
relativity are applicable in this metric. This is  not  the  case
with  the  original  string metric $g_{\mu\nu}$, which however is
the metric seen by the string. We  shall  confine
ourselves to the metric $\wt g_{\mu\nu}$.

The  model  can   be  extended  to have
electromagnetic  interactions by including the term
\begin{equation}
-{1\over 32\pi}\int d^4x\sqrt{-\wt g}e^{-2\phi}\wt g^{\mu\lambda}
\wt g^{\nu\rho}F_{\mu\nu}F_{\lambda\rho}
\end{equation}
in the action (\ref{2}). Exact black hole
solutions of this model have been found with non-zero charge and  angular
momentum.

The  black hole solution with zero angular momentum strongly resembles
the Schwarzschild solution of standard general relativity.
\begin{eqnarray}
d\wt s&=&\wt g_{\mu\nu}dx^{\mu}dx^{\nu}\nonumber\\
&=&-(1-{2M\over r})dt^2 +(1-{2M\over r})^{-1}dr^2 +r(r-a)d\Omega_
{II}^2\nonumber\\
e^{-2\phi}&=&e^{-2\phi_0}(1-{a\over r})\nonumber\\
F_{\theta\varphi}&=&Q\sin\theta
\label{dbh}\end{eqnarray}
where  $M$ is the mass of the black hole, $Q$ its charge, the parameter
$a$ is defined by
\begin{equation}
a={Q
^2\over 2M}e^{-2\phi_0}\label{a}\end{equation}
and $\phi_0$ is an arbitrary constant.
This black hole
has as usual a horizon at $r=2M$. An interesting feature
is that a curvature singularity occurs
at   $r=a$.  The  so-called  extremal solution corresponds to the
coincidence of these two regions and thus has $a=2M$. This
extremal limit is interesting also because the
area $4\pi .2M.(2M-a)$ of the horizon vanishes. All this is from the point of
view of the gravitational metric. However from the string theory
point of view  the geometry in  the  extremal
limit is perfectly non-singular. In the string metric the horizon
disappears  and  as  $r\to 2M$, the spacetime splits into a (1+1)
dimensional Minkowski spacetime times a sphere of constant radius
$2M$ (the {\it throat}).

We first wish to examine whether the area formula for entropy
is valid for this kind of black hole.
As argued in \cite{Hawk} the partition function for the
system can be defined by the (Euclidean) Lagrangian path integral for the
gravitational action coupled with matter fields. The dominant
contribution will come from the classical solutions of the action.
We may approximate the Euclidean action by taking something like
\begin{equation}
S_E[\wt g, \phi, F, \varphi]=
S_1[\wt g_{cl},\phi_{cl}, F_{cl}]+S_2[\wt g_{cl}, \varphi]+\cdots
\end{equation}
where $\varphi$ is the scalar field to be considered in the background
of the dilatonic black hole. Quantum fluctuations of the metric, the
electromagnetic field
and the dilatonic field are neglected and these variables are
frozen to their classical values. The
partition function can then be taken as
\begin{equation}
Z=e^{-S_1[\wt g_{cl}, \phi_{cl}, F_{cl}]}\int [d\phi]
e^{-S_2[\wt g_{cl}, \varphi]}.
\end{equation}

As in \cite{GH} the gravitational action has to be supplemented
by a surface term if the action is to be an extremum under variations
of the metric. The term has the form
\begin{equation}
S_{surface}={1\over 8\pi}\int d^3x \sqrt{-h}(K-K^0),\label{surf}
\end{equation}
where $h$ is the metric induced on the surface,
$K$ is the second fundamental form of the surface in the
metric $\wt g$ and $K^0$ is the same  in the flat metric
\begin{equation}
ds^2= -dt^2+dr^2+r^2d\Omega_{II}^2.
\end{equation}
The surface is taken to be  $S^1\times S^2$ where
$S^1$ has the circumference $\beta=8\pi M$ which can be
identified with the inverse temperature of the black hole. Then $K$ is
the covariant derivative of the unit vector
$n^r = (\wt g_{rr})^{-1/2}$ normal to the
surface:
\begin{eqnarray}
K\equiv -\nabla_r n^r&=&-n^r_{;\; r}-\Gamma^{\mu}_{r\mu}n^r\nonumber\\
&=&-(1-{2M\over r})^{1/2}{2r-3M-a(1-{M\over r})\over r(r-a)
}.\end{eqnarray}
Further, $ K^0=-2/r$.
If we consider a bounded region of large radius $r_0$ the surface term
(\ref{surf}) can be
evaluated using the solution (\ref{dbh}):
\begin{equation} S_{surface}={1\over 2}\beta (M+a)+{\cal O}({1\over r_0}).
\end{equation}
The electromagnetic contribution to the action for this solution
is given by
\begin{eqnarray}
S_{em}&=&-{1\over 32\pi}\int d^4x\sqrt{-\wt g}e^{-2\phi}\wt g^{\mu\lambda}
\wt g^{\nu\rho}F_{\mu\nu}F_{\lambda\rho}\nonumber\\&=&-{1\over 4}\beta a.
\end{eqnarray}
The contribution from the Einstein-Hilbert action is  not
identically zero in this case. Indeed, the
scalar curvature is given by
\begin{equation}
\wt R=-{a^2(1-{2M\over r})\over 2r^2(r-a)^2}.\end{equation}
But the kinetic term for dilatons  exactly
cancels the contribution from the Einstein- Hilbert action.
Finally the total action corresponding to this
solution is given by
\begin{equation} S_1[\wt g_{cl}, \phi_{cl}, F_{cl}]=
2\pi M(2M+a). \end{equation}

Following \cite{Hawk} we may replace $M$ everywhere by
$\beta/(8\pi)$. The free energy is then given by
\begin{eqnarray}
F&=&-{1\over\beta}\log Z\nonumber\\
&\approx& {\beta\over 16\pi}+{a(\beta)\over 4},
\end{eqnarray}
where $a(\beta)\propto 1/\beta$ (see (\ref{a})). Hence the entropy is
\begin{eqnarray}
S&=&\beta^2{\partial F\over\partial\beta}\nonumber\\
&=&{\beta^2\over 16\pi} - {a\beta\over 4}
\end{eqnarray}
which, by virtue of the relation  $\beta=8\pi M$,
is exactly equal to $Area/4$.
Thus the area formula is as much applicable for dilatonic black
holes as for ordinary black holes.

We come now to the contribution of the scalar field $\varphi$
to the partition function.
We   employ   the
brick-wall boundary condition \cite{'tHooft}. In this model the
wave function is cut off just outside the horizon. Mathematically,
\begin{equation}
\varphi(x)=0\qquad {\rm at}\;r=2M+\epsilon
\end{equation}
where   $\epsilon$  is a small, positive, quantity and signifies
an ultraviolet cut-off. There is also an infrared cut-off
\begin{equation}
\varphi(x)=0\qquad {\rm at}\;r=L
\end{equation}
with   $L>>2M$.

The  wave  equation  for  a scalar field in this
spacetime reads
\begin{equation}
\partial_{\mu}(\sqrt{-g}g^{\mu\nu}\partial_{\nu}\varphi)-m^2
\varphi=0.
\end{equation}
A solution of the form
\begin{equation}
\varphi=e^{-iEt}f_{El}Y_{lm_l}
\end{equation}
satisfies the radial equation
\begin{eqnarray}
(1-{2M\over r})^{-1}E^2f_{El}&+&{1\over r(r-a)}{\partial\over
\partial r}[(r-a)(r-2M){\partial f_{El}\over\partial r}]\nonumber\\
&-&[{l(l+1)\over r(r-a)}+m^2]f_{El}=0.
\end{eqnarray}
An $r$- dependent radial wave number can be introduced from  this
equation by
\begin{equation}
k^2(r,  l,  E)=  (1-{2M\over  r})^{-1}[(1-{2M\over r})^{-1} E^2 -
{l(l+1)\over r(r-a)} -m^2].
\end{equation}
Only such values of $E$ are to be considered here that the  above
expression  is  nonnegative. The values are further restricted by
the semiclassical quantization condition
\begin{equation}
n_r\pi=\int_{2M+\epsilon}^L~dr~k(r, l, E),
\end{equation}
where $n_r$ has to be a positive integer.

Accordingly, the free energy $F$ at inverse temperature $\beta$
is given by the formula
\begin{eqnarray}
\beta F&=&\sum_{n_r, l, m_l}\log(1-e^{-\beta E})\nonumber \\
&\approx  &  \int  dl~(2l+1)\int  dn_r\log   (1-e^{-\beta   E})
\nonumber\\
&=&-\int  dl~(2l+1)\int d(\beta E)~(e^{\beta E} -1)^{-1} n_r \nonumber\\
&=& -{\beta\over\pi}\int  dl~(2l+1)
\int dE~(e^{\beta E} -1)^{-1}\int_{2M+\epsilon}^L
dr~(1-{2M\over r})^{-1}\nonumber\\
&& \sqrt{E^2-(1-{2M\over r})({l(l+1)\over r(r-a)}+m^2)} \nonumber\\
&=& -{2\beta\over 3\pi}\int_{2M+\epsilon}^L dr~ (1-{2M\over r})^{-2}
r(r-a)\nonumber\\&& \int dE~(e^{\beta E} -1)^{-1}
[E^2-(1-{2M\over r})m^2]^{3/2}.
\end{eqnarray}
Here  the  limits  of  integration  for  $l, E$ are such that the
arguments  of  the  square  roots  are   nonnegative.   The   $l$
integration  is  straightforward  and has been explicitly carried
out. The $E$ integral can be evaluated only approximately.

The contribution to the $r$ integral from  large  values  of  $r$
yields   the  expression  for  the  free  energy  valid  in  flat
spacetime ($M=0$):
\begin{equation}
F_0=-{2\over 9\pi}L^3\int_m^\infty dE{(E^2-m^2)^{3/2} \over
e^{\beta E} -1}.
\end{equation}
We ignore this part \cite{GH, 'tHooft}.
The contribution of a  nonzero  $M$  is  singular  in  the  limit
$\epsilon\to 0$. The leading singularity is linear:
\begin{equation}
F_{lin}\approx -{2\pi^3\over  45\epsilon}(1-{a\over 2M})
({2M\over\beta})^4,
\end{equation}
where  the lower limit of the $E$ integral has been approximately
set equal to zero. If  the  proper  value  is  taken,  there  are
corrections  involving  $m^2\beta^2$  which will be ignored here.
This  result  reduces  to   the   formula \cite{'tHooft}  for   the
Schwarzschild  black hole when $a=0$. In general, there is simply
a multiplicative factor $(1-{a\over 2M})$.

There  is a logarithmic singularity as well, but it is in general
ignored because of the presence of the  linearly  divergent  term
$F_{lin}$.  However,  the  linear term vanishes when $a=2M$, {\it
i.e.}, when the black hole becomes extremal. In  this  case,  the
logarithmic term is the dominant one. It is
\begin{equation}
F_{log}\approx -{\pi^3\over  45M}\log({2M\over\epsilon})
({2M\over\beta})^4
\end{equation}
in the same approximation as above.

The entropy due to a nonzero $M$ can be obtained from the formula
\begin{equation}
S=\beta^2 {\partial F\over\partial\beta}.
\end{equation}
This gives
\begin{equation}
S=  {8\pi^3\over  45}({2M\over\beta})^3{(1-{a\over  2M}) (2M)^2
\over 2M\epsilon} {\rm for}~a\neq 2M
\end{equation}
and
\begin{equation}
S=  {8\pi^3\over  45}({2M\over\beta})^3 \log {(2M)^2
\over 2M\epsilon} {\rm for }~a=2M.
\end{equation}
Thus,  for  $a\neq  2M$,  namely  for nonextremal dilatonic black
holes, the Schwarzschild expression is valid, but with  the  area
factor   $(2M)^2$   corrected   by  the  appropriate  coefficient
$(1-{a\over 2M})$. Note that the factor $({2M\over\beta})$  is  a
constant  if  the  Hawking  temperature  is  used, because of its
inverse dependence on the mass, while the  quantity  $2M\epsilon$
may be regarded as giving an invariant measure of the distance of
the  brick  wall  from  the horizon \cite{'tHooft}. As we have shown
that the entropy of the dilatonic black hole itself
is   $S={\rm   (Area)}/(4G)$ (where $G$ has been set
equal to unity),
the   above   divergent
contribution  may  be  understood  as  a  renormalization  of the
gravitational  coupling  constant  $G$  \cite{Uglum}.  However,
quantum  gravity  being  non-renormalizable,  this interpretation
cannot  be  extended  to  include  quantum  fluctuations  of  the
gravitational fields.

In  the  case  of  {\it extremal}  dilatonic  black holes a logarithmic
formula  appears,  in  which  the  usual   factor   ${(2M)^2\over
2M\epsilon}$  is {\it replaced by its logarithm}. For these black
holes, where the area of the horizon  vanishes,  one  might  have
expected  the  entropy  to vanish altogether. What does happen is
that  the  linear  divergence  vanishes,  but   the   logarithmic
divergence,  which  is  of  course  weaker,  stays  on. A similar
logarithmic divergence  is  known  to  occur  if  the  theory  is
truncated to (1+1) dimensions \cite{Uglum}. Our calculation shows
that this is already present in (3+1) dimensions. But this divergence
cannot be regarded as a renormalization of $G$ because even with a
renormalized $G$, the zero area should have made the entropy zero.

To summarize, we have derived the area formula for the dilatonic
black hole from the Lagrangian path integral and
calculated semiclassically the entropy of a
scalar field in the background of this black  hole.  The
results  are  similar  to the case of ordinary black holes,
and involve a linearly divergent renormalization in general.
But the area  of  the
horizon may vanish here, and in that extremal case the
entropy of the scalar field does not vanish. The singularity becomes
logarithmic.

\end{document}